\documentclass[11pt]{article}
\usepackage[numbers]{natbib}
\usepackage{amsmath, amssymb}
\usepackage{graphicx}
\usepackage{booktabs}
\usepackage{hyperref}
\usepackage{natbib}
\usepackage{geometry}
\usepackage{booktabs}
\usepackage{float}

\title{AgroDesign: A Design-Aware Statistical Inference Framework for Agricultural Experiments in Python}

\author{Aqib Gul \\
IBPR, Sher-e-Kashmir University of Agricultural Sciences and Technology of Kashmir \\
\texttt{maaqib90@gmail.com}}

\date{}

\usepackage{graphicx}

\usepackage{caption}

\usepackage{float}   % optional but useful

\begin{document}

\maketitle

\begin{abstract}
Statistical analysis of agricultural experiments is based on structured experimental designs such as randomized block, factorial, split-plot, and multi-environment trials. While the theoretical bases of these approaches are sound, their implementation in modern programming frameworks usually involves manual specification of statistical models, choice of error terms, and subjective interpretation of interaction effects. This divide between experimental design and computational implementation opens the door to misleading inference and inconsistent reporting. We introduce AgroDesign, a Python framework that makes experimental design the central specification of statistical analysis. The framework translates specified experimental designs directly into valid linear models, automatically identifies error strata, conducts hypothesis testing and mean separation, checks assumptions of linear models, and provides decision-focused interpretations. The framework integrates fixed-effect ANOVA, hierarchical designs, linear mixed models, and genotype-by-environment stability analysis into a single declarative framework. AgroDesign is validated on canonical designs in agricultural statistics and shows consistency with traditional statistical analysis while strictly enforcing correct interpretation constraints, especially in interaction-dominant and multi-stratum designs. By integrating design semantics into computation, the framework minimizes analyst-driven modeling choices and enhances reproducibility.
\end{abstract}

\section{Introduction}

\subsection{Statistical analysis in agricultural experimentation}

Controlled experimentation is the basis of inference in agricultural and biological sciences. Field and greenhouse experiments are often carried out using organized experimental layouts such as completely randomized designs (CRD), randomized complete block designs (RCBD), factorial designs, split-plot experiments, and multi-environment trials. Such designs provide the means for distinguishing between treatment and environmental variations and are the basis for statistically valid agronomic recommendations \citep{Gomez1984}.
In contrast to general regression analysis, the inference in designed experiments relies not only on the linear model but also on the design of randomization and the corresponding error strata (Morris, 2010). In blocked and hierarchical designs, each factor needs to be compared against the proper experimental unit. The wrong mean square value leads to improper F-tests and, consequently, to potentially incorrect scientific inferences \citep{Steel1997,Piepho2003}. Therefore, a correct analysis involves the determination of the experimental unit, the setup of the proper analysis of variance table, the comparison of treatments, the validation of model assumptions, and the interpretation of statistical results in agronomic terms \citep{Gomez1984}. Although the theoretical basis of ANOVA is firmly grounded, its proper application demands a high level of expertise in both statistical modeling and experimental design theory. This often leads to the dependence of the validity of conclusions on the analyst's expertise rather than on the documented design of the experiment.
With the growing influence of computational workflows on scientific analysis, the lack of a formal representation of experimental structure introduces a disconnect between the theory of experimental design and its implementation in software \citep{Garijo2014}. The same data set can lead to different statistical inferences based on the structure of the model formulation, despite the experimental design being the same. This shows that experimental design is more than a conceptual framework for planning and analysis and is instead a form of structured information that needs to be explicitly encoded in the analytical process. Experimental design as a first-class computational object offers a means to encode the structure of randomization, restrict model formulation, and ensure that statistical inference is based on the principles of classical analysis of variance and not on analyst interpretation.

\subsection{Limitations of current computational workflows}

Contemporary environments for scientific computing support rich statistical modeling but mostly in regression-centric frameworks. For instance, Python packages like statsmodels support linear modeling and ANOVA with formula interfaces \citep{Seabold2010}. Although mathematically self-contained, these packages require the analyst to explicitly specify interaction terms and corresponding error structures. This task is especially prone to errors in hierarchical experimental designs like split-plot or multi-factor studies. Other packages for scientific computing offer discrete analytical functionalities. Numerical computing toolkits offer hypothesis tests and distributions \citep{Virtanen2020}, and specialized packages offer multiple comparison procedures (Terpilowski, 2019). Experimental design matrix generators offer designs but lack statistical inference. Hence, the process of analysis involves combining disparate tools, making analysis more complex and error-prone.
Reproducibility studies have also found that analytical flexibility and hidden researcher choices can affect statistical results \citep{Stodden2013,Sandve2013}. In the case of designed experiments, manual model building and post-hoc testing introduce further degrees of freedom, making it hard to guarantee consistent interpretation across analysts and experiments. There are domain-specific statistical packages for designed agricultural experiments in other ecosystems, such as specialized statistical platforms and packages in the R environment. However, these packages are often isolated from contemporary data science workflows used for preprocessing, data visualization, and machine learning. This results in redundant effort and prevents integration with reproducible computational research workflows.
In agricultural statistics, there has been a need for specialized software to handle this issue, and this has been done in design-oriented analysis environments. For instance, there are specialized software packages like agricolae in R that have procedures for treatment comparison and analysis in classical experimental designs \citep{Mendiburu2021}. However, these software packages work as standalone statistical software and not as part of the general computational workflow. Modern scientific computing is now incorporating data processing, visualization, and machine learning into a unified programming environment, but the analysis of experimental design is done outside this environment. This means that scientists have to work between environments, exporting data for statistical analysis and then importing the results back for further processing.

\subsection{Contribution of this work}

This research presents a design-driven framework for statistical inference, where the experimental design becomes the central specification of analysis instead of a statistical formula specified by the user. The experimental design is specified by the analyst, and the framework automatically derives the statistical procedures needed for inference. Based on the experimental design, the system automatically builds the corresponding analysis of variance model, identifies the correct error strata, separates treatment means, checks assumptions for the model, and provides statistically meaningful results. The analysis, therefore, follows directly from the encoded randomization design instead of analyst-specified modeling choices.
The framework is implemented in an open-source Python system called agrodesign, which considers experimental design as a first-class computational object. Instead of being a set of loosely coupled statistical tests, the system imposes a strict inference workflow that connects model building, hypothesis testing, post-hoc testing, and biological interpretation.
The contributions of this work are:
\begin{enumerate}
\item A declarative framework for experimental design that captures experimental units and randomization hierarchy as part of the computational process.
\item Automatic inference of linear model structure and corresponding error terms based on the specified design.
\item A constrained inference process that combines ANOVA, multiple testing, and assumption checks to minimize analyst-specified parameters.
\item A decision component that converts statistical outcomes to treatment comparisons and agronomic interpretations.
\item Empirical verification by comparison with standard statistical methods on simulated and experimental data.
\end{enumerate}
By formalizing experimental structure as an executable analytical specification, the proposed framework enables reproducible and statistically consistent analysis of agricultural experiments while remaining compatible with modern scientific computing environments.

\section{Statistical Framework}

\subsection{Experimental design as a structured statistical model}

In classical analysis of variance, the statistical model is not specified independently of the experiment but is specified by the randomization scheme used to create the data. Every experimental design implicitly specifies (i) the set of estimable effects, (ii) the appropriate experimental unit for each effect, and (iii) the denominator mean square for valid hypothesis testing \citep{Gomez1984, Montgomery2017}. In this case, the experimental design can be considered as a structured statistical object rather than descriptive metadata.
Let Y be the response variable measured on experimental units created through a randomization procedure. Design specifies a hierarchy of units $U_1 \subseteq U_2 \subseteq \cdots \subseteq U_k$, on which treatments are applied at certain levels of the hierarchy. The validity of inference relies on comparing each treatment effect to variability at the same level of randomization. Thus, two experiments with the same data but different assignment structures are different and specify different statistical models.
For example, in a completely randomized design with treatment factor $T_i$, observations satisfy
\begin{equation}
Y_{ij} = \mu + \tau_i + \varepsilon_{ij}, \quad 
\varepsilon_{ij} \sim \mathcal{N}(0, \sigma^2)
\end{equation}
where all experimental units share a common error variance.
In a randomized complete block design, an additional block factor $B_j$ represents environmental heterogeneity:
\begin{equation}
Y_{ij} = \mu + \tau_i + B_j + \varepsilon_{ij}
\end{equation}
where treatment effects are tested against within-block variability rather than total variability. Hierarchical designs introduce multiple error strata. In a split-plot experiment with whole-plot factor A and sub-plot factor B, two independent error terms exist:
\begin{equation}
Y_{ijk} = \mu + A_i + B_j + (AB)_{ij} + n_{k(i)} + \varepsilon_{ijk}
\end{equation}
where $n_{k(i)}$ represents whole-plot error and $e_{ijk}$ represents subplot error. Testing a factor against $e_{ijk}$ instead of $n_{k(i)}$ produces incorrect F-statistics (Steel et al., 1997).
Similarly, multi-environment trials introduce crossed random factors:
\begin{equation}
Y_{ijk} = \mu + G_i + E_j + (GE)_{ij} + \varepsilon_{ijk}
\end{equation}
where interpretation depends on the magnitude and significance of the interaction term \citep{Piepho2003}.
These examples show that the design of the experiment shapes the statistical model, the tests of hypotheses, and the region of interpretation. The use of design information as external annotation implies that the analyst must manually infer these constraints, which can be ambiguous and prone to errors. The approach outlined in this paper formalizes experimental design as a computational representation of this structure. Rather than specifying statistical models, the design specification captures the nesting of experimental units and the assignment of treatment. The statistical model and valid inference procedures are then inferred from this representation.

\subsection{Model construction from experimental structure}

Based on a given data set and a stated experimental design, the framework identifies an admissible statistical analysis that corresponds to the models described in Section 3.1. The experimental design not only specifies which linear model is to be fitted but also specifies which estimator and hypothesis tests are appropriate. In this way, estimation and inference are decoupled, and parameters are estimated in the standard statistical manner, while the design specifies the set of permissible inferential operations.
For single-level experimental units, the stated design specifies the completely randomized model (1). Comparisons of treatments are based on residual variation. With blocks stated, the randomized complete block model (2) is specified, and comparisons of treatments are made within blocks rather than over the entire data set. For factorial designs, the crossed model (3) is automatically specified, and all interaction terms specified by the design are included without analyst request.
Hierarchical layouts bring in multiple experimental units. Split-plot designs with models of type (3) estimate the full linear model, but test hypotheses by error strata defined for each factor. Effects for whole plots are tested against whole-plot error, while subplot effects are tested against the residual error. The estimation method stays the same, but the validity of inference is defined by the randomization scheme, not by the linear model. Some experiments contain both fixed effects for treatments and random effects for environmental factors. When random factors are specified, the analysis imposes a linear mixed model of the form
\setcounter{equation}{4}  % So next equation becomes (5)
\begin{equation}
Y = X\beta + Z u + \varepsilon
\label{eq:mixed}
\end{equation}
where fixed effects measure treatment performance and random effects measure environmental variation. Rather than comparing simple means, treatment performance is assessed through best linear unbiased predictors, yielding adjusted estimates that consider different environments and a lack of replication. The goal of inference shifts from comparing means to predicting treatment performance in a random environment.
For multi-environment trials modeled by equation (4), the genotype-by-environment interaction defines the range of valid inferences. The specified interaction structure is then further decomposed to describe stability and adaptability. Additive main-effect and multiplicative interaction decomposition, Finlay and Wilkinson regression, and Eberhart and Russell stability parameters describe both mean performance and consistency of response to environments. These inferences are not considered distinct analyses but rather implications of the same underlying model, thereby maintaining statistical consistency between stability analysis and performance prediction.
The same structural model is generalized over multiple response variables and grouped experiments. For multiple traits, the same design-based model is specified independently for each response while maintaining the same structure for interpretation.
For grouped datasets such as repeated years or locations, the dataset is partitioned into subsets
\begin{equation}
D = \bigcup_{g \in G} D_g
\label{eq:union}
\end{equation}
and the same model implied by the experimental design is applied within each subset. The resulting analyses form a structured collection of comparable inference results because the statistical procedure remains invariant across groups.

\subsection{Hierarchical inference protocol}

Statistical significance is not sufficient to draw valid conclusions in designed experiments. The question of drawing conclusions based on treatment effects is related to the hierarchical structure of the model fitted to the data. In factorial and hierarchical designs, lower-order effects may not be meaningful if higher-order interactions are present. However, standard statistical software packages provide all the tests at once, and the analyst is left to conclude, which may not be consistent.
The approach presents a constrained inference problem based on the structure of the fitted model. Let the fitted model include a set of effects
\begin{equation}
\mathcal{F} = \{E_1, E_2, \dots, E_n\}
\label{eq:effectset}
\end{equation}
each associated with a hypothesis test. Define the order of an effect as the number of factors involved in that term. The admissible interpretation domain is determined by the highest-order statistically significant effect
\begin{equation}
E^* = \arg\max_{E_i \in \mathcal{F}}
\left\{
\operatorname{order}(E_i)
\mid
p(E_i) \le \alpha
\right\}
\label{eq:dominanteffect}
\end{equation}
The admissible interpretation domain is determined by the dominant effect defined in Equation~\ref{eq:dominanteffect}. All lower-order effects that do not conform to $E^*$ are excluded from interpretation. Therefore, in the case of a significant interaction, marginal means are not to be interpreted since they are averages of responses to treatments that are not homogeneous. Rather, inference is limited to simple effects in the joint factors. Post-hoc analysis is carried out only within the admissible region. Let M(E) denote the estimated marginal means for the effect E. Multiple comparison procedures are restricted to
\begin{equation}
\mathcal{M} =
\left\{
M(E)
\mid
E \in E^*
\right\}
\label{eq:meanset}
\end{equation}
Therefore, comparisons are made among treatment combinations when interactions are dominant and among main effects only if no higher-order interaction exists. This avoids making potentially invalid comparisons based on averaging over non-uniform treatment responses.
The above hierarchical restriction can be generalized to other factorial models. In split-plot designs, inferences are restricted to factors defined over different units of experiments. In mixed models, treatment comparisons are made based on predicted performance rather than raw values because of the random environmental effects. In multi-environment trials, overall recommendations or environment-specific recommendations are valid depending on the size of the genotype by environment interaction. By incorporating the rules of inference directly into the inference process, the analysis becomes deterministic based on the experimental design and significance level. The analyst defines the experiment, and the valid conclusions are derived based on the structure of the model.

\subsection{Assumption validity and diagnostic constraints}

The inferential steps outlined above are based on distributional assumptions related to linear models. For hypothesis testing and treatment comparison to be valid, the residuals of the model must meet the assumptions under which the F-statistic and multiple comparison procedures are derived. It is therefore important to note that assumption checking is considered part of the inferential process and not a diagnostic step after analysis.
Let $\hat{\varepsilon}_{ijk}$ denote residuals from the fitted model. For fixed-effect analyses, admissible inference requires
\begin{equation}
\hat{\varepsilon}_{ijk} \sim \text{i.i.d. } \mathcal{N}(0, \sigma^2)
\label{eq:residualassumption}
\end{equation}
which implies two practical conditions: residual normality and homogeneity of variance across treatment groups. Normality is necessary for the validity of the approximation of the F-distribution in hypothesis testing, while homoscedasticity is necessary for the mean squares in the denominators to be unbiased estimators of error variance. Violations do not prevent estimation but condition the validity of inference. The framework, therefore, introduces a validity predicate
\begin{equation}
\mathcal{V}(D, \mathcal{E}) =
\begin{cases}
\text{true}  & \text{if model assumptions are not rejected at level } \alpha_v, \\
\text{false} & \text{otherwise.}
\end{cases}
\label{eq:validity}
\end{equation}
Hypothesis tests and treatment comparisons are interpreted conditionally on $\mathcal{V}$. When the assumptions are met, conclusions are considered statistically valid. Assumption checks are evaluated at the error stratum corresponding to each tested effect.
In hierarchical and mixed models, the same holds for the residual part related to each experimental unit. For split-plot designs, assumption checks are relevant to the subplot residuals related to the error term employed in hypothesis testing. For mixed models, assumption checks are relevant to the conditional residuals after adjusting for random effects. In this manner, the validity criterion is contingent upon the error stratum that determines inference rather than the entire data set. By integrating assumption checks into the inference process, statistical conclusions are made contingent upon model validity rather than implicitly accepted. The analytical process thus makes a distinction between calculable outcomes and valid conclusions, ensuring that conclusions based on the experiment are consistent with the probabilistic assumptions that underlie the statistical inference.

\subsection{ Decision-oriented interpretation}

The last step of the framework translates statistically acceptable conclusions into a formal decision rule. In designed experiments, statistical significance is not the criterion for practical conclusions but rather relies on the objective of the experiment and the formal structure of the effects inferred. The framework views statistical output as an intermediate form rather than the result of analysis.
Let $\mathcal{M}$ denote the admissible set of estimated means determined in Section~2.3. 
A decision function $\delta : \mathcal{M} \to \mathbb{R}$ maps statistically valid comparisons to a recommendation set $\mathbb{R}$. The function is defined according to the type of model implied by the experimental design.
In fixed-effect experiments, the decision rule is based on ranking treatments by estimated marginal means, considering the grouping constraints imposed by multiple comparison tests. Treatments belonging to the same statistical group are considered equivalent, and the recommendation is stated in terms of a set of statistically equivalent optimal treatments, not a single numerical maximum. In situations where interaction effects are dominant, ranking is conditionally made within factor combinations rather than across marginal averages. In mixed-model analysis, decisions are made based on predicted performance rather than observed means. Let $\hat{u}$ represent predicted treatment effects derived from the mixed model. The recommendation rule chooses treatments that maximize predicted response, considering random environmental variation, such that the decision is based on expected performance rather than observed outcomes.

\section{Implementation}

\begin{figure}[H]
    \centering
    \includegraphics[width=0.85\textwidth]{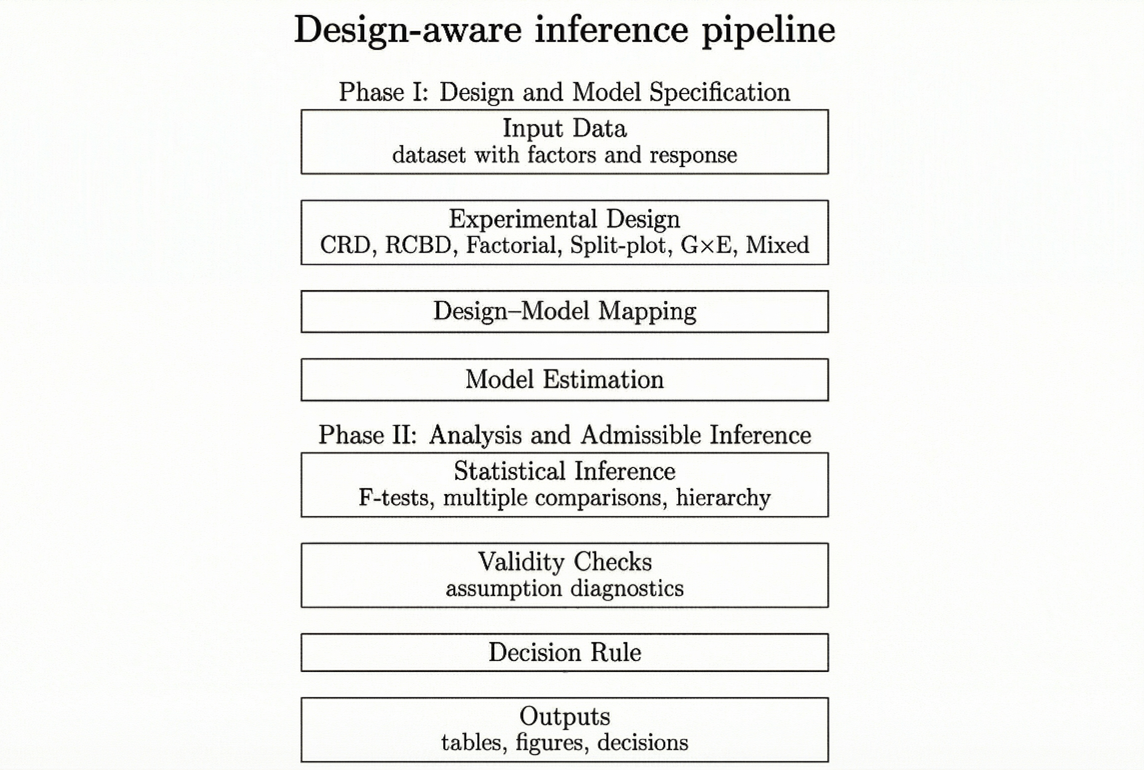}
    \caption{
    Design-aware inference pipeline implemented in \texttt{agrodesign}. 
    Phase I maps experimental design to the appropriate statistical model. 
    Phase II performs admissible inference, assumption validation, and decision generation.
    }
    \label{fig:pipeline}
\end{figure}

The framework is implemented as a computational system that executes the statistical procedure defined in Section 2. Rather than introducing new estimators, the implementation enforces correct model construction and admissible inference while delegating numerical computations to established statistical libraries. The software, therefore, acts as an execution layer for the statistical framework. The overall architecture of the framework is illustrated in Figure~\ref{fig:pipeline}. The pipeline is divided into two stages: design specification and admissible inference.
During the specification phase, the experimental design is stated and mapped to a corresponding statistical model. The system builds up the correct model structure according to the experimental design, factor hierarchy, and experimental units, and carries out parameter estimation using proven numerical algorithms. This step identifies the estimable effects but does not yet conclude.
During the inference phase, statistical tests are assessed using the proper error strata corresponding to each factor. Multiple comparison procedures and hierarchical interpretation rules limit the valid set of comparisons. Model validity is then checked using diagnostic tests, and only statistically valid conclusions are passed on to the decision rule. The result is a set of structured conclusions based on the constrained inference space rather than statistical tables. The experimental design thus acts as an executable specification of analysis. The design of the implementation carefully distinguishes between statistical methodology and numerical computation. Parameter estimation is based on proven statistical algorithms, while the framework manages model structure, inference validity, and interpretation logic. This distinction maintains numerical accuracy while ensuring that statistical inferences are consistent with the experimental design.

\section{Experimental Validation}

All validation examples demonstrate agreement with classical experimental design theory. To ensure that the proposed framework is producing statistically valid inference and not simply replicating software results, analyses were conducted on a set of controlled agricultural experiments that represent different experimental structures. Each dataset represents a canonical design class that is commonly used in agricultural research: completely randomized experiments, blocked trials, factorial treatments, hierarchical field designs, and multi-environment breeding trials.
Validation focuses on four criteria:
\begin{enumerate}
\item Correct identification of the statistical model and denominator error.
\item Correct restriction of interpretation domain.
\item Consistent mean separation results.
\item Coherent agronomic recommendation derived from admissible inference.
\end{enumerate}
Rather than evaluating predictive accuracy, the objective is methodological correctness: the framework should reproduce conclusions a trained statistician would obtain when applying classical experimental design theory.

\subsection{Completely Randomized Design (CRD)}

\begin{figure}[H]
    \centering
    \includegraphics[width=0.8\textwidth]{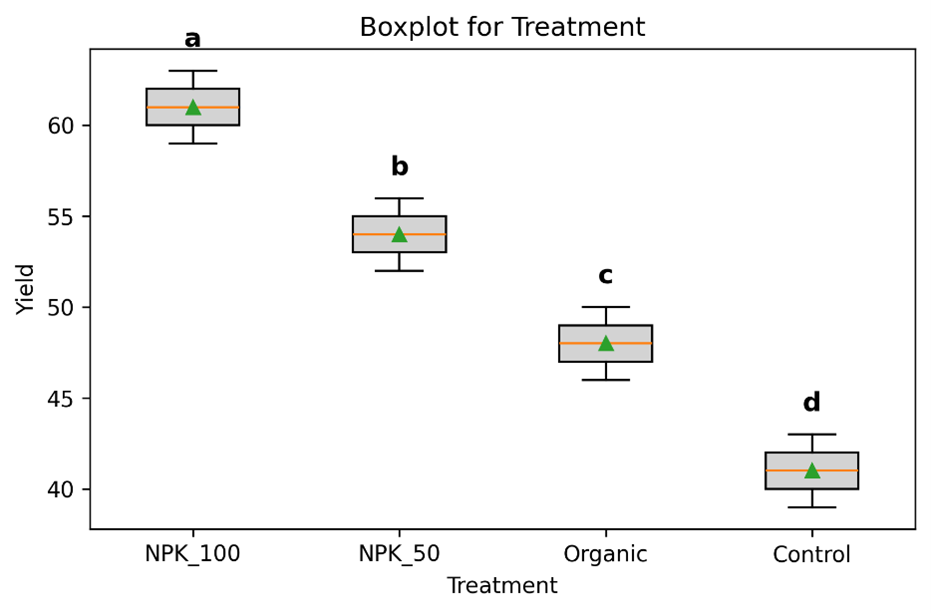}
    \caption{Variety comparison and compact letter display for the RCBD experiment. 
    Letters denote Tukey HSD groupings at $\alpha = 0.05$ after adjusting for block effects.}
    \label{fig:crd_boxplot}
\end{figure}

The first validation involves a completely randomized experiment where treatments are administered independently to similar experimental units. In this experiment, the observations are described by model (1), where the treatment effects are contrasted with a common residual variance. The model automatically builds the fixed-effect model and the analysis of variance table given in Table~\ref{tab:crd_anova}
\begin{table}[htbp]
\centering
\caption{Automatically generated ANOVA for the CRD experiment.}
\label{tab:crd_anova}
\begin{tabular}{lcccc}
\hline
Source & DF & MS & F & p-value \\
\hline
Treatment & 3  & 363.333 & 145.333 & $<0.001$ \\
Residual  & 16 & 2.500   & --      & -- \\
\hline
\end{tabular}
\end{table}
The treatment effect is highly significant (F = 145.33, p < 0.001), which confirms the existence of yield differences among fertilizer treatments. Since the design includes a single factor without any hierarchical structure, the admissible interpretation domain includes only the treatment main effect. Mean comparison based on Tukey HSD indicates that there are four distinct groups, and the highest mean yield is produced by NPK100. Since there is no interaction structure, there are no restrictions on treatment comparison. The treatment comparison is illustrated in Figure~\ref{fig:crd_boxplot}.
The graph verifies the mean separation test result. Treatments with different letters have non-overlapping confidence intervals, and the median order corresponds to the ANOVA test result. The graph shows that the framework enables the creation of publication-quality agronomic analyses directly from the decision layer of the statistical model. Residual plots are automatically produced. The Shapiro-Wilk test shows a slight violation of normality (p-value = 0.034), illustrating that the framework correctly states assumption violations without changing the model structure or decision logic. 

\subsection{Blocking validation (Randomized Complete Block Design (RCBD))}

\begin{figure}[H]
    \centering
    \includegraphics[width=0.8\textwidth]{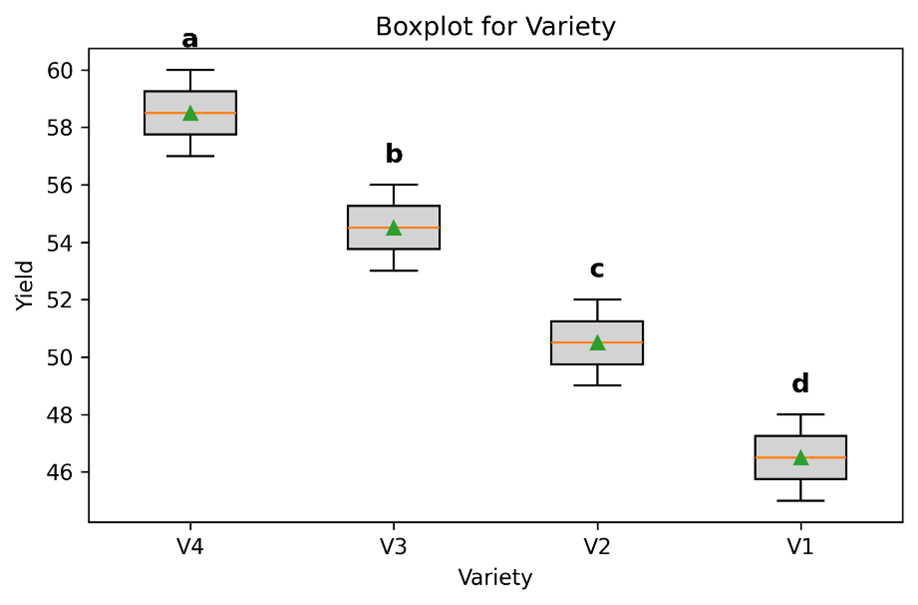}
    \caption{
    Variety comparison after blocking adjustment in the RCBD experiment. 
    Letters denote Tukey HSD groupings at $\alpha = 0.05$. 
    Mean differences reflect within-block contrasts.
    }
    \label{fig:rcbd_boxplot}
\end{figure}

To check whether the framework is properly addressing the issue of blocking structure, a dataset from a randomized complete block design (RCBD) was used. In the RCBD, the comparison of treatments needs to be assessed after removing the variability between blocks, and hence, the treatment variance is compared against the residual variance within blocks as shown in Table~\ref{tab:rcbd_anova}.
\begin{table}[htbp]
\centering
\caption{Automatically generated ANOVA for the RCBD experiment.}
\label{tab:rcbd_anova}
\begin{tabular}{lcccc}
\toprule
Source   & DF & MS      & F       & p-value \\
\midrule
Variety  & 3  & 106.667 & 320.000 & $<0.001$ \\
Block    & 3  & 5.667   & 17.000  & $<0.001$ \\
Residual & 9  & 0.333   & --      & -- \\
\bottomrule
\end{tabular}
\end{table}
The treatment effect is highly significant (F = 320.00, p < 0.001), while block effects are also significant, confirming the presence of environmental heterogeneity across blocks. Mean separation identifies four distinct varietal groups, with variety V4 producing the highest yield. The varietal separation is illustrated in Figure~\ref{fig:rcbd_boxplot}.
After adjusting for block variation, the differences between varieties are still well distinguished. The graphical display captures the comparison of treatments based on the appropriate error term, demonstrating that the nuisance variability has been accounted for before making comparisons among treatments. Since there is no interaction structure in this experimental design, the analysis stops at the main effect level after adjusting for block variation. The residual plot shows that there is no deviation from normality (Shapiro-Wilk test p-value = 0.064), justifying the validity of the model fit.

\subsection{Factorial experiment (interaction-aware inference)}

\begin{figure}[H]
    \centering
    \includegraphics[width=0.85\textwidth]{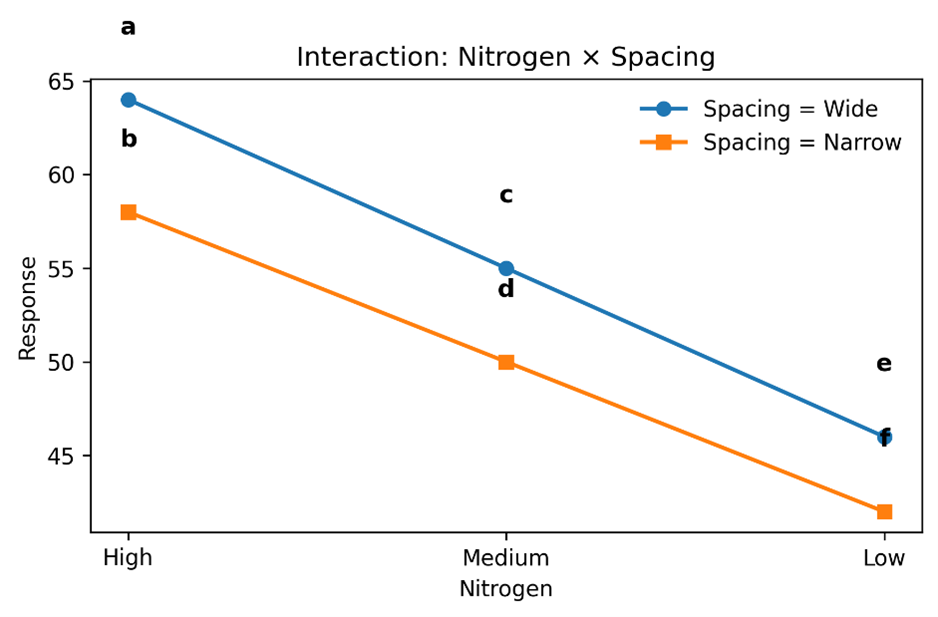}
    \caption{
    Nitrogen $\times$ Spacing interaction plot for the factorial experiment. 
    Nearly parallel response profiles indicate absence of interaction, 
    supporting marginal interpretation of main effects.
    }
    \label{fig:factorial_interaction}
\end{figure}

To test the automatic construction of multi-factor models, a two-factor factorial dataset with nitrogen and plant spacing was examined. In factorial experiments, both main effects and interaction effects are considered, and results are based on the highest-order significant effect. The ANOVA results for the factorial experiment are presented in Table~\ref{tab:factorial_anova}. 
\begin{table}[htbp]
\centering
\caption{Automatically generated ANOVA for the factorial experiment.}
\label{tab:factorial_anova}
\begin{tabular}{lcccc}
\toprule
Source & DF & MS & F & p-value \\
\midrule
Nitrogen & 2 & 433.500 & 433.500 & $<0.001$ \\
Spacing & 1 & 112.500 & 112.500 & $<0.001$ \\
Nitrogen $\times$ Spacing & 2 & 1.500 & 1.500 & 0.262 \\
Residual & 12 & 1.000 & -- & -- \\
\bottomrule
\end{tabular}
\end{table}
Both nitrogen and spacing main effects are highly significant, whereas the interaction term is not (F = 1.50, p = 0.262). Because the interaction is not statistically significant, the admissible interpretation domain consists of the two main effects. 
The similarity in response trends suggests the absence of interaction (Figure~\ref{fig:factorial_interaction}, as confirmed by the ANOVA result that marginal interpretation is valid. 
The framework thus allows independent main-effect interpretations. Mean separation suggests three different levels of nitrogen and two levels of spacing. As no interaction is dominant, ranking of treatments is done using marginal means. The maximum yield is obtained from the treatment High × Wide, but as there is no significant interaction, the result is justified based on independent main-effect interpretation. The residual analysis suggests a departure from normality (Shapiro-Wilk p-value = 0.0016).

\subsection{Split-plot experiment (multi-stratum inference validation)}

\begin{figure}[H]
    \centering
    \includegraphics[width=0.85\textwidth]{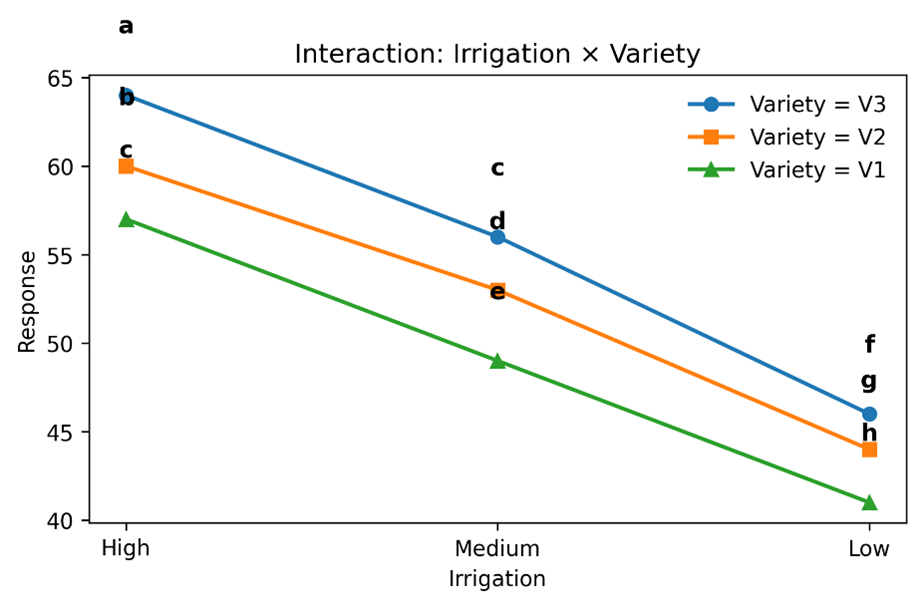}
    \caption{
    Irrigation $\times$ Variety interaction in the split-plot experiment. 
    Non-parallel response profiles indicate interaction-dominant inference, 
    requiring interpretation at the combination level.
    }
    \label{fig:splitplot_interaction}
\end{figure}

Split-plot designs contain multiple experimental units and therefore require different denominator mean squares for different factors. Whole-plot factors must be tested against the whole-plot error term, while subplot factors and their interactions are tested against the subplot residual error. Misidentification of these strata is a common source of incorrect inference in applied research. The ANOVA results for the split-plot experiment are presented in Table~\ref{tab:splitplot_anova}. The interaction between irrigation and variety is statistically significant, whereas the whole-plot error term (Block $\times$ Irrigation) is not.
\begin{table}[htbp]
\centering
\caption{Automatically generated ANOVA for the split-plot experiment.}
\label{tab:splitplot_anova}
\begin{tabular}{lcccc}
\toprule
Source & DF & MS & F & p-value \\
\midrule
Block & 2 & 8.111 & 73.0 & $<0.001$ \\
Irrigation & 2 & 626.333 & 5637.0 & $<0.001$ \\
Variety & 2 & 90.333 & 813.0 & $<0.001$ \\
Block $\times$ Irrigation & 4 & 0.111 & 1.0 & 0.445 \\
Irrigation $\times$ Variety & 4 & 1.333 & 12.0 & $<0.001$ \\
Residual & 12 & 0.111 & -- & -- \\
\bottomrule
\end{tabular}
\end{table}
The irrigation $\times$ variety interaction is highly significant 
($F = 12.0$, $p < 0.001$), indicating that varietal performance 
depends on irrigation level. Consequently, interpretation is restricted 
to treatment combinations rather than marginal main effects. 
This confirms that the framework correctly prioritizes the highest-order 
significant effect in determining the admissible interpretation domain.
The interaction structure is illustrated in 
Figure~\ref{fig:splitplot_interaction}. 
The non-parallel response lines confirm that varietal ranking 
changes across irrigation levels, supporting interaction-dominant inference.
A large part of the variability comes from random block effects, thus justifying variance-adjusted treatment ranking. This illustrates the automatic transition from fixed-effect comparison to variance-component inference. Treatment T3 has the largest adjusted performance after adjusting for block variability. Since treatment ranking is done using predicted values and not marginal means, the recommendation is variance-adjusted treatment effects. 

\subsection{Linear mixed model validation}

\begin{figure}[H]
    \centering
    \includegraphics[width=0.65\textwidth]{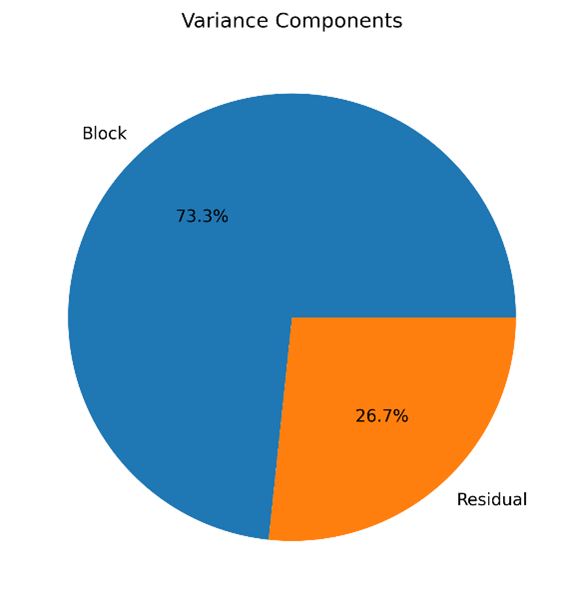}
    \caption{
    Proportion of total variance attributable to block and residual components 
    under the linear mixed model.
    }
    \label{fig:lmm_variance}
\end{figure}

Some agricultural experiments need to model the environment or blocking effects as random rather than fixed. To test the variance component estimation and treatment ranking, a mixed effect data set was analyzed with blocks as random effects and treatments as fixed effects. The automatically estimated variance components are shown in Table~\ref{tab:lmm_variance}.
\begin{table}[htbp]
\centering
\caption{Estimated variance components from the linear mixed model.}
\label{tab:lmm_variance}
\begin{tabular}{lc}
\toprule
Component & Variance \\
\midrule
Block (random) & 1.22 \\
Residual (random) & 0.44 \\
\bottomrule
\end{tabular}
\end{table}
The positive block variance confirms the presence of environmental heterogeneity, justifying the mixed-model specification. Treatment performance is evaluated using best linear unbiased predictors (BLUPs), which adjust treatment means for random block variation. BLUP-based treatment performance is summarized in Table~\ref{tab:lmm_blup}. 
\begin{table}[htbp]
\centering
\caption{BLUP-based treatment ranking under the mixed model.}
\label{tab:lmm_blup}
\begin{tabular}{lc}
\toprule
Treatment & BLUP \\
\midrule
T3 & 54.5 \\
T2 & 48.5 \\
T1 & 41.5 \\
\bottomrule
\end{tabular}
\end{table}
Figure~\ref{fig:lmm_variance} illustrates the relative contribution 
of block and residual variance components.
The results for the mixed models show that the ranking of treatments is based on estimates of variance, rather than on the raw marginal means. The incorporation of the effects of blocks as random in the model changes the inference from fixed effect ANOVA to BLUP inference without changing the interface. This result is consistent with the expectation that the system combines traditional variance component analysis with design-informed decision logic. The ability to switch between fixed-effect and mixed-effect 
inference within a unified design specification highlights 
the flexibility of the framework.

\subsection{Multi-environment trial (G×E) validation}

To evaluate performance across environments, a genotype-by-environment (G×E) dataset was analyzed. In such experiments, conclusions depend on the magnitude of the interaction term: strong interaction implies environment-specific recommendations, whereas weak interaction permits global ranking. The ANOVA results for the multi-environment trial are presented in 
Table~\ref{tab:gxe_anova}. 
\begin{table}[htbp]
\centering
\caption{Automatically generated ANOVA for the multi-environment trial.}
\label{tab:gxe_anova}
\begin{tabular}{lcccc}
\toprule
Source & DF & MS & F & p-value \\
\midrule
Genotype & 3 & 364.458 & 728.917 & $<0.001$ \\
Environment & 3 & 42.458 & 84.917 & $<0.001$ \\
Genotype $\times$ Environment & 9 & 0.458 & 0.917 & 0.535 \\
Residual & 16 & 0.500 & -- & -- \\
\bottomrule
\end{tabular}
\end{table}
Genotype $\times$ environment effects are significant, while the interaction term is not (p = 0.535). Therefore, treatment ranking is valid across environments, and marginal genotype performance can be interpreted. The framework accordingly performs BLUP-based ranking (Table~\ref{tab:gxe_blup}):
\begin{table}[htbp]
\centering
\caption{Genotype BLUP-based ranking across environments.}
\label{tab:gxe_blup}
\begin{tabular}{lc}
\toprule
Genotype & BLUP \\
\midrule
G4 & 7.7 \\
G3 & 2.7 \\
G2 & -2.3 \\
G1 & -8.1 \\
\bottomrule
\end{tabular}
\end{table}

\begin{figure}[H]
    \centering
    \includegraphics[width=0.75\textwidth]{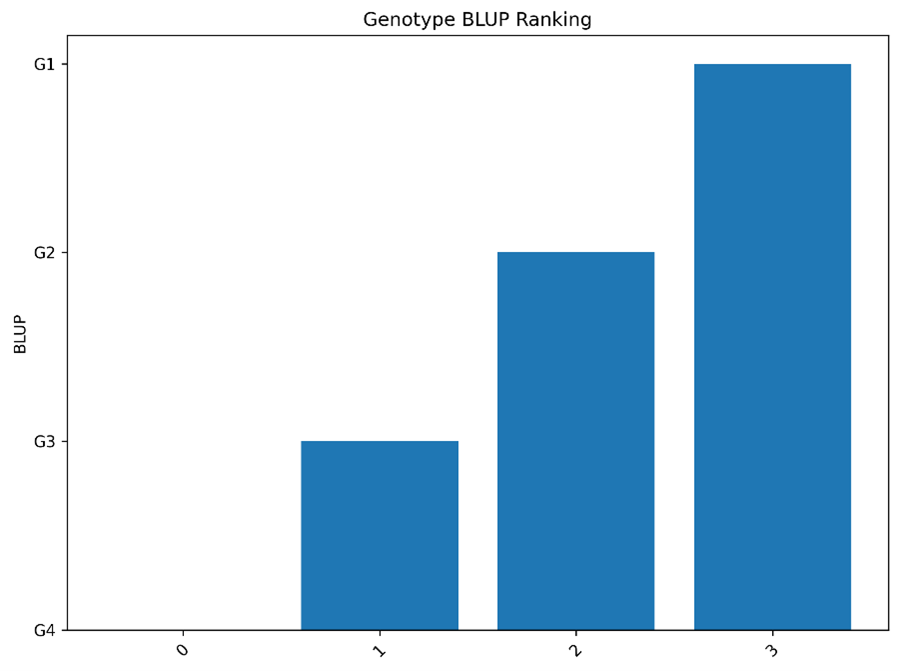}
    \caption{
    BLUP-based genotype ranking across environments. 
    Positive values indicate above-average performance.
    }
    \label{fig:blup_ranking}
\end{figure}

Figure~\ref{fig:blup_ranking} provides a graphical representation 
of BLUP-based genotype performance. 

\begin{figure}[H]
    \centering
    \includegraphics[width=0.85\textwidth]{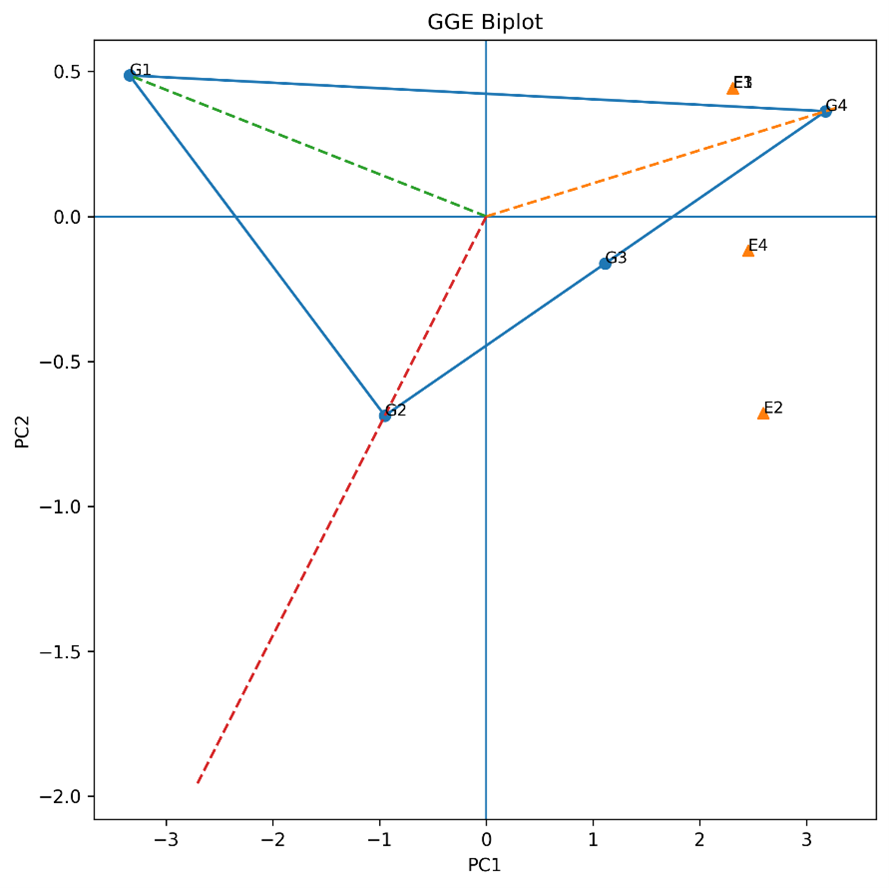}
    \caption{
    GGE biplot of genotype-by-environment performance. 
    Proximity to the average environment axis indicates wide adaptation.
    }
    \label{fig:gge_biplot}
\end{figure}

The predicted performance across environments reveals the superiority of genotype G4, as expected by the non-significant interaction. The genotype G4 has the highest predicted performance. The broad-sense heritability is estimated as $H^2 = 0.99$, which reflects strong genetic control and the effectiveness of selection. As the interaction is non-significant, the stability analysis confirms the wide adaptation instead of superiority in individual environments. 
The biplot (Figure~\ref{fig:gge_biplot}) shows the existence of a single mega-environment and stable genotype performance, which agrees with the statistical result on wide adaptation and justifies the decision rule extracted from the ANOVA structure. The model thus proposes a global genotype instead of environment-specific recommendations. The residual analysis shows normality (p = 0.121), which justifies the validity of the model. 

\section{Discussion}

The paper enunciates a fundamental principle of experimental design: the relationship between statistical inference and the randomization framework of the experiment \citep{Gomez1984, Steel1997}. While classical theory establishes this relationship, contemporary computational analysis often requires this relationship to be manually incorporated through regression analysis. AgroDesign closes this gap by incorporating design semantics into the analysis workflow.
One of the most important aspects is the enforcement of hierarchical interpretation rules. In factorial and split-plot designs, the effects of lower orders may be uninterpretable in the presence of higher-order interaction effects \citep{Montgomery2017}. In standard statistical software, all effects are calculated and reported simultaneously. The current system formalizes this rule as a deterministic computational constraint: the permissible interpretation region is automatically reduced to the highest-order significant effect. As a result, statistical results are made structurally interpretable rather than numerically descriptive. The system also combines fixed-effect and mixed-model inference in a single design-aware framework. Agricultural experiments often involve variation in environmental conditions that necessitate variance component estimation and prediction, as opposed to mean comparison. With the current system, factors can be designated as random using the same interface, and the workflow will automatically shift from ANOVA-based inference to BLUP-based estimation, as in mixed-model analysis \citep{Piepho2003}. This obviates the requirement for analysts to change paradigms or software platforms when transitioning from field experiments to breeding trials.
Multi-environment analysis illustrates an additional benefit: the direct mapping of statistical interaction structure to decision logic. Whether or not genotype-by-environment interaction is present determines whether global advice or environment-specific advice is valid. Stability methods such as AMMI and regression-based methods are typically conducted as separate analyses. In the current system, they become outcomes of the same structural model, ensuring that stability analysis and treatment ranking are logically consistent. From a software development perspective, the main innovation is not the provision of new statistical estimators but the encoding of experimental design as an executable specification. Instead of choosing tests based on inspection of the data, the researcher specifies the experimental design, and the possible statistical manipulations follow deterministically. This improves specification uniqueness, especially in hierarchical experiments where wrong choice of denominator is still a common mistake in applied practice. The system also promotes computational reproducibility. Flexibility in analysis and hidden modeling assumptions are acknowledged causes of irreproducible scientific results \citep{Stodden2013}. By imposing a rule-governed workflow on model building, interpretation, and comparison, the system eliminates analyst-specific variability while maintaining compatibility with general Python data science workflows.
However, several limitations exist. Validation is presently restricted to balanced canonical designs and should be extended to unbalanced and missing-data field trials. Assumption checks are at present limited to normality and homogeneity of variance. Future research could generalize the framework to generalized linear and Bayesian models. Finally, automated interpretation is not a substitute for agronomic knowledge but rather provides a statistically consistent foundation upon which expert knowledge can act. In summary, this research illustrates that representing experimental design as a computational object is a way to bring statistical theory into alignment with software development. The system developed here mandates proper inference, combines classical ANOVA and mixed models, and provides reproducible agronomic advice in a contemporary computing environment.

\section{Conclusion}

This research presented a design-driven framework for statistical inference, where the experimental design serves as the central specification for analysis. Rather than forcing the analyst to specify statistical models by hand, the system automatically infers the admissible structure for the model, tests of hypotheses, and rules for interpretation based solely on the stated randomization structure of the experiment. This framework combines fixed-effect ANOVA, hierarchical designs, mixed models, and multi-environment stability analysis into a single declarative analysis pipeline. By incorporating hierarchical interpretation constraints and proper error strata identification into the inference process, the system ensures that all stated conclusions are consistent with known principles of experimental design. The software thus changes statistical analysis from a series of analyst choices to a deterministic transformation from experimental structure to admissible inference and recommendation. Experimental validation on traditional agricultural designs shows that the system correctly reproduces the conclusions that would be reached by classical statistical analysis while eliminating the specification ambiguity. The system's outputs, such as treatment comparison and agronomic recommendations, are based on statistically valid domains of comparison rather than model outputs. Incorporating design semantics into executable computation enhances reproducibility and reduces variability. Future development will extend support to generalized response distributions, unbalanced field trials, and larger multi-location breeding datasets. Nevertheless, the present framework establishes a foundation for reproducible, design-aware statistical analysis in Python, enabling experimental conclusions to follow directly from the experiment performed.

\section{Software Availability and Reproducibility}

AgroDesign is distributed as an open-source Python package and is publicly available through the Python Package Index (PyPI). The software can be installed using:

\begin{verbatim}
pip install agrodesign
\end{verbatim}

The full source code, documentation, version history, and issue tracker are maintained in a public GitHub repository:

\noindent\textbf{Repository:} \\
\url{https://github.com/DeepStatistix/AgroDesign}

All datasets used for experimental validation are included within the package as built-in tutorial datasets representing canonical experimental designs, including completely randomized designs (CRD), randomized complete block designs (RCBD), factorial experiments, split-plot designs, linear mixed models, and multi-environment trials (G$\times$E). These datasets are synthetic and are provided exclusively for methodological verification and reproducibility.

Each analysis presented in this article can be reproduced programmatically using the package's built-in dataset loader and default workflow. The framework automatically generates analysis of variance tables, variance component estimates, multiple comparison outputs, diagnostic tests, graphical visualizations, and treatment ranking summaries.

Because model specification is derived directly from the declared experimental design, the analytical pipeline is deterministic. Consequently, all statistical outputs reported in this paper correspond exactly to executable code in the public repository. The article documents the methodological and structural principles underlying the framework, while the repository provides full computational reproducibility.

\clearpage
\bibliographystyle{plainnat}

\clearpage
\appendix
\section{Minimal Reproducible Example}

This appendix demonstrates a complete analysis workflow using the
\texttt{AgroDesign} framework. The example illustrates how a randomized
complete block design (RCBD) experiment can be analyzed from raw data
to agronomic recommendation using a single declarative specification.

\subsection*{Installation}

The package is publicly available on the Python Package Index (PyPI)
and can be installed using:

\begin{verbatim}
pip install agrodesign
\end{verbatim}

\subsection*{Loading a Built-in Dataset}

AgroDesign provides built-in datasets representing canonical agricultural
experimental designs. The following example uses a variety trial arranged
in a randomized complete block design.

\begin{verbatim}
from agrodesign.datasets import load_dataset
from agrodesign.experiment import Experiment

df = load_dataset("rcbd")
df.head()
\end{verbatim}

The dataset contains yield observations for multiple crop varieties
evaluated across blocks representing field heterogeneity.

\subsection*{Running the Analysis}

The user specifies the experimental structure rather than a statistical model:

\begin{verbatim}
result = Experiment(df, "Yield").rcbd("Variety", "Block").run()
\end{verbatim}

The framework automatically determines the appropriate linear model,
constructs the analysis of variance table, and applies the correct
error term based on the declared design.

\subsection*{Statistical Report}

A full statistical report can be printed directly:

\begin{verbatim}
print(result)
\end{verbatim}

This command automatically generates:

\begin{itemize}
    \item Analysis of variance (ANOVA) table
    \item Multiple comparison results
    \item Interpretation hierarchy
    \item Assumption diagnostics
\end{itemize}

\subsection*{Agronomic Recommendation}

The agronomic interpretation can be obtained independently of the
statistical report:

\begin{verbatim}
result.summary()
\end{verbatim}

This produces a concise decision-oriented summary consistent with
the admissible interpretation domain.

\subsection*{Visualization}

Publication-ready figures are generated automatically:

\begin{verbatim}
result.plot()
\end{verbatim}

\subsection*{Exporting Reproducible Outputs}

All tables and figures can be exported to a structured results directory:

\begin{verbatim}
result.export("results")
\end{verbatim}

The exported folder contains:
\begin{itemize}
    \item Formatted ANOVA tables
    \item Treatment comparison summaries
    \item Diagnostic plots
    \item Complete textual report
\end{itemize}
\end{document}